# Van der Waals Ferromagnetic Josephson Junctions

Linfeng Ai[1,2,†], Enze Zhang[1,2,†], Ce Huang[1,2,†], Xiaoyi Xie[1,2], Yunkun Yang[1,2], Zehao Jia[1,2], Yuda Zhang[1,2], Shanshan Liu[1,2], Zihan Li[1,2], Pengliang Leng[1,2], Xingdan Sun[3], Xufeng Kou[4], Zheng Han[3], Faxian Xiu[1,2,5,6*]

[1] State Key Laboratory of Surface Physics and Department of Physics, Fudan University, Shanghai 200433, China
[2] Institute for Nanoelectronic Devices and Quantum Computing, Fudan University, Shanghai 200433, China
[3] Shenyang National Laboratory for Materials Science, Institute of Metal Research, Chinese Academy of Sciences, Shenyang 110016, China
[4] School of Information Science and Technology, ShanghaiTech University, Shanghai 201210, China
[5] Collaborative Innovation Center of Advanced Microstructures, Nanjing 210093, China
[6] Shanghai Research Center for Quantum Sciences, Shanghai 201315, China

[†]These authors contributed equally to this work.
[*]Correspondence and requests for materials should be addressed to F. X. (E-mail: Faxian@fudan.edu.cn).

**Abstract:**

**Superconductor-ferromagnet (S-F) interfaces in two-dimensional (2D) heterostructures present a unique opportunity to study the interplay between superconductivity and ferromagnetism. The realization of such nanoscale heterostructures in van der Waals (vdW) crystals remains largely unexplored due to the challenge of making an atomically-sharp interface from their layered structures. Here, we build a vdW ferromagnetic Josephson junction (JJ) by inserting a few-layer ferromagnetic insulator $Cr_2Ge_2Te_6$ into two layers of superconductor $NbSe_2$. Owing to the remanent magnetic moment of the barrier, the critical current and the corresponding junction resistance exhibit a hysteretic and oscillatory behavior against in-plane magnetic fields, manifesting itself as a strong Josephson coupling state. Through the control of this hysteresis, we can effectively trace the magnetic properties of atomic $Cr_2Ge_2Te_6$ in response to the external magnetic field. Also, we observe a central minimum of critical current in some thick JJ devices, evidencing the coexistence of 0 and $\pi$ phase coupling in the junction region. Our study paves the way to exploring the sensitive probes of weak magnetism and multifunctional building blocks for phase-related superconducting circuits with the use of vdW heterostructures.**

**Introduction**

Since the discovery of graphene[1], two-dimensional (2D) van der Waals (vdW) crystals have been widely investigated on their heterostructures[2,3] by integrating disparate materials, promoting the exploration of unusual phenomena such as charge transfer[4] and proximity effect[5–7]. By taking advantage of excluding the dangling bonds and lattice mismatch at surfaces, high-performance nanodevices have also been demonstrated in electronics[8] and spintronics[9]. Among these, the discovery of intrinsic layered superconductors[10] (S) and ferromagnets[11–13] (F) whose properties vary in layer numbers paves the way to further exploiting such an incompatible order-parameter system. As such, the design of S-F heterostructures mainly focuses on the proximity-effect-dominant coupling like Josephson junctions (JJ) with ferromagnetic barriers. With atomically flat and magnetically sharp interfaces, this type of functional nanodevices may potentially serve as critical components for superconducting quantum circuits[14] or memories[15].

In a ferromagnetic JJ, exchange energy ($E_{ex}$) in magnets imposes a finite momentum to Cooper pairs, leading to a modulation of spatially-damped oscillations of superconducting order parameter when penetrating the ferromagnetic barrier over the decay length scale[16]. Depending on the JJ barrier thickness, the ground state can either be a 0-junction with equal superconducting phases on both sides or a π-junction where the phases differ by π. Consistent with theoretical predictions, experiments have shown the presence of 0-π transitions by tuning the F-layer thickness in metallic or alloyed magnets[17,18], as well as 0-π JJ structures[19] with a corporate area consisting of these two ground states, providing the potential for an arbitrary φ-phase control[20] in superconducting electronics. Except for the mentioned above, magnetic insulators in JJs are also fascinating for their spin filtering tunnel behaviors[21], in which a pure second-harmonic current phase relation[22] and incomplete 0-π transitions[23] have been realized. Also, inspired by the recent progress in the graphene-based micromagnetometry[24], the JJ technique can be analogically developed to probe the weak magnetism of vdW magnetic insulator, because of its sensitivity to spin-dependent Josephson tunneling process when applied as a tunnel barrier.

Here we report the construction of 2D-vdW JJs with ferromagnetic insulating barriers. Layered dichalcogenide 2H-NbSe$_2$ is an excellent vdW superconductor for its robust superconductivity[25], which has previously been selected to form highly transparent JJs[26,27]. For the other, we choose intrinsic ferromagnetic semiconductor Cr$_2$Ge$_2$Te$_6$ with a bulk Curie temperature ($T_{Curie}$) of ~ 61 K[28] as an intermediate layer, whose easy axis points out of the cleaved plane. The JJ shows a typical Fraunhofer pattern manifesting a strong Josephson coupling state. Besides, the observed hysteresis of critical current ($I_c$) along with the corresponding junction resistance ($R$) against field sweep ascribes to the barrier's magnetism. The measured hysteresis depends on the maximum applied field ($B_{max}$), implying an evolutionary magnetization process of the barrier remanent magnetic moment. Meanwhile, the hysteretic behavior only appears when the JJs are in the superconducting state. We also find a central minimum of $I_c$ patterns in some thick barrier devices as a typical 0-π JJ signature, evidencing the coexistence of both 0 and π phase meditated coupling in the whole junction region. Interestingly, the multidomain state in few-layer $Cr_2Ge_2Te_6$ can introduce an asymmetric field-dependent in-plane magnetization, resulting in an asymmetric supercurrent distribution in the two halves of 0 and π phase segments.

**Results**
**vdW JJs based on 2D S/F materials**

Ferromagnetic JJ devices were fabricated by a dry-transfer technique[29] (see Methods), which produces a clean interface via vdW coupling between layers. Figure 1a and 1b are top and side views of the layered structures of $NbSe_2$ and $Cr_2Ge_2Te_6$, respectively. Figure 1c is a schematic illustration of the device structure with a four-terminal measurement configuration, where 2H-$NbSe_2$ flakes are placed on the top and bottom and few-layer $Cr_2Ge_2Te_6$ acts as an insulating barrier. The whole stack is fabricated on a silicon substrate capping of 285 nm $SiO_2$. We choose a laminated vertical junction structure which is suitable for a relatively small piece of exfoliated $Cr_2Ge_2Te_6$. Figure 1d (left) displays a false-color scanning electron microscope (SEM) image of device #01 with trilayers labeled by the dashed lines, and the height profile of intermediate $Cr_2Ge_2Te_6$ flake is around 6.5 ± 0.4 nm measured by the atomic force microscope (AFM), beneath a flat junction area of ~ 14.2 $\mu m^2$. The entire fabrication process is performed in a glove box. A carefully controlled procedure assisted by the stepper motor is applied to eliminate the bubbles and surface oxidation, making it a high-quality vdW interface.

The JJ characteristics were investigated below the superconducting critical temperature ($T_c$) of the $NbSe_2$ crystals. Figure 1d (right) shows typical current-voltage (*IV*) curves of junction device #01 at 0.1 K under zero field, while tiny differences in the trap and retrap current situations reveal a weak-link type junction with small capacitance described by Resistively-Shunted-Junction (RSJ) model[30]. In Figure 1e, a contour plot of junction differential resistance (*dV/dI*) as a function of bias current (*I*) and in-plane field ($B_{//}$) is presented. Periodic oscillations of $I_c$ from the Fraunhofer pattern correspond well to its in-plane *R*, as the positions of $I_c$ minima are the same as those of *R* maxima. Also, clear second-order oscillations are seen at higher bias due to Fiske resonance[31] when the *a.c.* Josephson frequency equals the eigenfrequency of electromagnetic oscillations in this cavity, proving a well-defined JJ through the vdW stacking. The periodicity of $I_c$ oscillation *ΔB* ~ 4.2 mT is given by an integer flux quantum $\Phi_0$ in the effective area ($S_{eff}$), from which we obtain the effective junction length $L_{eff} = S_{eff} / W$ = 79 nm (*W* ~ 6.2 μm is the junction width perpendicular to the field direction). The Josephson penetration depth $\lambda_J = [h/4\pi\mu_0 e L_{eff} J_c]^{1/2}$ is estimated to be 30.6 μm at 0.1 K which is much larger than the junction width and length, where *h* is Planck's constant, $\mu_0$ is the vacuum permeability, *e* is the electron charge and $J_c$ (~ 353.5 A/$cm^2$) is the critical current density. Thus our devices should be treated as being in the short junction limit.

For an unsaturated magnetic barrier, it is feasible to continuously reverse its orientations at a lower magnetic field. In Figure 2, we present a comparison of Fraunhofer oscillation patterns of device #01 (~ 6.5 nm) when the field starts from different $B_{max}$. Initially, it was magnetized to 9 mT at 0.1 K, then descended from positive to negative fields at -9 mT and eventually returned. The separated sweep branches in Figure 2a give the central maximal $I_c$ offsets at -0.6 mT and 0.3 mT, respectively. The total flux enclosed in the junction is the sum of the external flux and additional flux generated by the barrier magnetic moment, the latter of which reverses smoothly at a low magnetic field, resulting in a hysteresis window of 0.9 mT. The shape of patterns is slightly distorted under the magnetized process, as the field widths of the first side-lobes of the pattern are not exactly the same as those at opposite field directions, consistent with the model of a field-dependent magnetic moment in a hysteresis loop. The hysteresis observed in zero-bias $R$ is equivalent to $I_c$ (shown in Figure 2b). Next, we magnetized the device to 15 mT and repeated the same measurement at the higher applied field. Figure 2c shows that the positions of zero-flux $I_c$ shift to -2 mT and 0.8 mT, resulting in a larger window of 2.8 mT. Estimations from $I_c$ and $R$ measurements in Figure 2d show accordant sizes of hysteresis. Nevertheless, the values measured at the negative magnetic field are smaller than those at the positive field, suggesting an asymmetric supercurrent distribution from the two nonequivalent S/F hetero-interfaces vertically.

To further understand the evolution of barrier magnetization, first, we obtain its hysteretic in-plane $R$ with a $B_{max}$ of 100 mT at 0.1 K (a log-plot of Figure 3a (up) is used to clarify zero $R$), in which the flux-induced oscillations in $R$ still persist when the supercurrent is unmeasurable at higher field. The coincidence over ±75 mT of these two sweep directions reveals a saturation field of barrier magnetization. Empirically, we can correspond each minimum/maximum of $R$ to integer/half-integer flux quantum trapped in the junction[32] (Figure 3a (down)). Therefore, the magnetization ($M$) of the barrier can be reconstructed from the equation $\Phi = 4\pi M S_{eff} + \Phi_H$, where $\Phi$ is the sum-up flux embedded in a JJ and $\Phi_H$ is generated by the external field. Meanwhile, the largest hysteresis ($\Delta B_{max}$), estimated from the central position of zero-order minimum $R$, can also be regarded as a quantitative index for this nanoflake's in-plane coercivity (in device #01, $\Delta B_{max}$ is 9.3 mT when fully saturated to 100 mT). Next, we show the plot of $\Delta B_{max}$ varying with $B_{max}$ at 0.1 K in Figure 3b, from which we find that $\Delta B_{max}$ has a positive correlation with $B_{max}$ and nearly converges above 90 mT.

We also investigated the hysteretic behavior with thermal excitation, conjointly resolving the features about how hysteresis vanishes above the superconducting $T_c$. Figure 3c depicts the hysteretic in-plane $R$ of the same device compared between the superconducting state and normal state. The hysteresis shrinks in $R$ amplitude and magnetic field range as temperature increases and finally disappears above $T_c$ since it is non-superconducting, which seems contradictory to the fact that $Cr_2Ge_2Te_6$ at 6.5 K is still in the ferromagnetic state as the hysteresis originates from its remanence. For detailed analyses, we measured the variation of $\Delta B_{max}$ as a function of temperature with a $B_{max}$ of 27 mT along with the normalized $R$ under zero field (Figure 3d). The value of $\Delta B_{max}$ remains almost unchanged below 2 K and then gradually decreases when warming up to 5.5 K. Nevertheless, the discrepancies in the curves of 6 and 6.2 K turn to become illegible around zero field causing a sudden drop of $\Delta B_{max}$ (see

Supplementary Figure S4). Therefore, we assume that $\Delta B_{max}$ here can denote the magnitude of Josephson interference. The reduction of $\Delta B_{max}$ is predominantly determined by thermal fluctuation above 2 K, suppressing the phase coherence between two separated superconductors. This assumption is further supported by the bias-dependent experiments. A set of hysteretic $R$ loops at *d.c.* bias current varying from 0 to 200 μA is shown in Figure 3e, from which $\Delta B_{max}$ was obtained when the junction was driven into several superconducting states (qualitatively distinguished in the 0 mT $dV/dI$ curve). We can observe an identical $\Delta B_{max}$ around 3.3 mT (within the error bar) below 100 μA, in line with our expectations that $\Delta B_{max}$ is only influenced by a specific $B_{max}$ before saturation at a specific temperature. Besides, the deviation between two directional sweeps is practically unidentifiable at a high bias of 200 μA and 400 μA (see Supplementary Figure S5) in good agreements with a relatively small $\Delta B_{max}$ measured above 6 K, as a consequence of the main contribution to the junction $R$ from the normal state $R$ of superconductor $NbSe_2$.

**Evidences of π phase coupling**

For those spin-singlet pairing superconductors, the presence of a π-JJ requires a comparable F-layer thickness to the oscillation wavelength as the schematic views shown in Figure 4a and 4b. The Josephson phase is correlated through the whole junction even in the case that 0 and π phase supercurrents coexist in the same region, leading to a 0-π JJ structure. As the spontaneous flux penetrations into these two parts produce a counterproductive effect, they will conjointly result in a central minimum of $I_c$ instead of a maximum value either in absolute 0-JJs or π-JJs. Here, we find similar signatures of 0-π JJs in thick barrier devices which may also be the evidence of a φ-phase state. Figure 4c is the $I_c$ oscillation patterns of device #02 in the range of 30 mT, whose barrier thickness is around $10.5 \pm 0.3$ nm. Extracted $I_c$ values at a positive bias ($I_c^+$) are given in Figure 4d and the hysteresis with regard to the opposite sweeping direction is around 3 mT, acquired from the positions of central minimum of $I_c$. The pronounced characteristics also perform in another device #03 whose barrier thickness is of the same magnitude of around $11 \pm 0.5$ nm, giving the hysteresis of around 2 mT with a $B_{max}$ of 50 mT as shown in Figure 4e and 4f. Both two devices have a comparatively $I_c$ background due to the finite-voltage criterion effect. Besides, the two sweeping branch curves of $I_c$ coincide above 30 mT, implying a smaller in-plane coercivity of a thicker $Cr_2Ge_2Te_6$ than that of a thinner one measured in device #01. It should be mentioned that the influence of Abrikosov vortex[33] from superconductor electrodes trapped in the JJ also needs to be considered, for which it can introduce anomalous phase shifts leading to a similar shape of $I_c$ patterns. There are two criteria to distinguish, one is the vortex-induced hysteresis that is opposite to remanent magnetization[34,35], which is contradictory to our observations. The other can be inferred from the symmetry of its interference pattern. For a conventional JJ, $I_c$ should be symmetric with respect to $I_c^+(B) = I_c^-(B)$ and $I_c(+B) = I_c(-B)$. Distorted $I_c$ patterns by the vortex will naturally accord with the former formula, yet neither of them has been satisfied in our devices results. Therefore we can attribute this 0-π phase signature to be mediated by a magnetic layer.

Deviated from the symmetric 0-π junction model, we find remarkable asymmetries of the $I_c$ patterns in our devices. First, non-zero central minimum of $I_c$ can be originated from the asymmetry in 0 and π phase critical current densities. Besides, unequal $I_c$ values and lobe widths of the two halves of subsequent peaks around the central valley should ascribe to different field-dependent flux penetration to 0 and π parts, introduced by the barrier in-plane magnetization[19]. Last, the non-standard periodic dependence of $I_c$ oscillations compared between the positive and negative field applied halves indicates a strong modulation by inhomogeneous local magnetizations. The nature of a possible multi-domain state in few-layer $Cr_2Ge_2Te_6$[36] is expected to generate such asymmetries in this ferromagnetic barrier. The pinning of the domains can be reconfigured under the external in-plane magnetic field in the two halves, therein resulting in the observed variations of the central minimum as well as successive maximum of $I_c$ values with regard to the opposite field sweeping directions.

**Discussion**

Our experiments on ferromagnetic JJs via vdW connection demonstrate a hysteretic supercurrent and resistance behavior, dominated by the barrier magnetization related to the $B_{max}$ and temperature. We would refocus on the validity of using the JJ structure to probe magnetism in atomic vdW materials. Few experiments have shown a direct study of ferromagnetic semiconductors or insulators via transport measurements, particularly limited at low temperatures. Like previously reported structures of tunneling magnetoresistance (TMR)[37] or spin-orbit torque (SOT)[38] devices for the same goal, the JJ is also a potential tool for its high sensitivity to minor variations in the response of external magnetic fields, because of a dramatic transition between the normal and superconducting states. The thought of employing the flux quantization as an *in-situ* magnetometer proves it to be a precise technique to trace the evolution process of its barrier magnetization.

Since the unsaturated barrier moment can reverse smoothly at low fields, the supercurrent here follows a continuous modulation by the external flux shown in $I_c$ measurements above. However, with progressive magnetization to a higher field, the Josephson tunneling process tends to become fragile, as the emergence of breakpoints in $I_c$ causes a discontinuous pattern (see Supplementary Figure S7 and S8). In device #01, the discontinuity starts from 23 mT, with an evident signature of abrupt changes of $I_c$ between two flanks of a breakpoint. Another feature is the irreversibility of the whole pattern as the breakpoints persist when the field sweeps back. Detailed characterizations are performed in the Supplementary Materials. This picture might be associated with the layered spin structures of $Cr_2Ge_2Te_6$, accounting for its enhanced interlayer coupling at high fields, that the barrier magnetization will resist external changes, thus a larger step of the applied field is required to overcome such a potential. When the fully-aligned spins start to flip back, the spin interactions between separated layers prevent it from a smooth reverse leading to the observed irreversibility. Besides, pinned spins of the ferromagnetic layer adjacent to the superconducting layer may also be restricted due to their 2D circumstances, strongly modulated by impurities and obstacles at two heterointerfaces.

The multi-domain structure of the barrier is also expected to generate various effects in the diffusive JJs. For a thinner barrier, the temperature dependence of $I_cR_n$ product ($V_c$) of device #01 (~ 6.5 nm) is a well fit to Ambegaokar-Baratoff (AB) relation[39] for an insulating barrier in the dirty limit. The calculated zero temperature $V_{c0}$ ~ 0.51 mV is smaller than the theoretical maximum value of $\pi\Delta/2e$ = 1.57 mV ($\Delta$ ~ 1 meV is the gap value of bulk NbSe$_2$), showing a strong suppression by the magnetic barrier on Josephson current for the phase differences acquired by spin-up or spin-down electrons. Compared to the suppression, we find an apparent excess supercurrent over the AB relation fit at low temperature (see Supplementary Figure S6) in a thicker Josephson device #03 (~ 11 nm). The $I_cR_n$ product gained at 0.5 K exceeds the diffusive limit from the AB relation fit. The gradual deviation between the data and AB values below 4.6 K is similar to former studies in NbN/GdN/NbN JJs [ref. 21]. Considering our material systems, the equilibrium domain size of Cr$_2$Ge$_2$Te$_6$ is probably larger in ultrathin nanoflakes as mediated by a single-domain remanence, in contrast to the tendency of forming multi-domains when approaching a 3D bulk limit. Aware of the discrepancy in barrier thickness, the relative size of domain walls would introduce a notable magnetic inhomogeneity that implies the reduction of exchange energy seen by the Cooper pairs, therein promoting the enhancement of Josephson current in thicker samples[40].

To conclude, we report the realizations of Josephson junctions based on 2D van der Waals superconductor NbSe$_2$ and ferromagnet Cr$_2$Ge$_2$Te$_6$. We demonstrate that such hysteretic behaviors of supercurrent and magnetoresistance are predominantly induced by the magnetic barrier remanence. Moreover, we observe 0-π JJs signatures which are the evidence of π phase coupling in this crystalline vdW heterostructure. The layered structure of the barrier can generate an anomalous tunneling effect for its internal interlayer coupling. The stacked nanojunctions are of great potential to investigate vdW coupling in layers, and further expectations on dissipationless spin-active or switchable φ-phase JJs can be accomplished in the use of other insulating magnets like CrI$_3$ or CrCl$_3$ with intrinsic anisotropic magnetization.

**Note:** During the preparation of this manuscript we became aware of a recent preprint[41] with similar materials/device structures, and with thorough discussions and theoretical calculations to probe and engineer the JJ phase. Compared to their results, we mainly focus on thicker barriers in JJ devices and show the behavior of Fraunhofer patterns under magnetic fields while they demonstrate the Ising Cooper pairs tunneling through a multi-domain ultra-thin Cr$_2$Ge$_2$Te$_6$ layer.

**Methods**
**Crystal growth.** High-quality crystal of bulk NbSe$_2$ was grown via chemical vapor transport method with iodine as the transport agent. A stoichiometric ratio of Nb and Se powders (with 0.2 % excess of Se) were evacuated and sealed in a quartz tube with 0.1 g iodine flakes, and then placed in a two-zone furnace in a temperature gradient from 730 °C to 770 °C for two weeks. Single-crystals of Cr$_2$Ge$_2$Te$_6$ were synthesized using the Te self-flux method. The raw powders of Cr, Ge and Te (with a ratio of 1 : 4 : 20) were mixed and kept at 950 °C for 6 hours, and the mixture was then cooled at a rate of 2 °C/h, followed by a centrifugation at 500 °C to get its bulk material.

**Device fabrication.** The heterostructure devices were fabricated through a dry-transfer method in the glove box ($H_2O$, $O_2$ < 0.1 ppm). Few-layered $NbSe_2$ and $Cr_2Ge_2Te_6$ flakes were first mechanically exfoliated onto polydimethylsiloxane (PDMS) stamps with rough identifications of thickness by optical microscope, and next stacked onto a Si substrate capped with a 285 nm-thick $SiO_2$ layer slowly, controlled by a stepper motor. The multi-terminal electrical contacts were patterned by a standard *e*-beam lithography technique and subsequently deposited of Cr/Au (5 / 70 nm) via magnetron sputtering.

**Transport measurements.** Four-terminal transport measurements were carried out in two Physical Property Measurement Systems (PPMS, Quantum Design), one equipped with a dilution refrigerator with the temperature down to 50 mK and the other without down to 1.9 K. The transport properties were acquired using lock-in amplifiers (SR 830) and Agilent 2912 meters. In differential resistance measurements, a small *a.c.* excitation was generated from the lock-in amplifier output voltage in combination with a larger *d.c.* bias applied by Agilent 2912 through a 100 kΩ resistor, and the differential voltage was measured at a low frequency (< 40 Hz). In magneto-transport measurements, the field direction was parallel to the substrate, therefore perpendicular to the current flowing across the barrier from top to bottom superconductors.

**Ambegaokar–Baratoff theory.** AB theory[39] is an exact result for the full temperature dependence of critical current, expressed as $I_c(T)R_n = (\pi\Delta(T)/2e)\cdot tanh[\Delta(T)/2k_BT]$, where $\Delta(T)$ is the superconducting gap calculated from BCS theory.

**Acknowledgments**
This work was supported by the National Key Research and Development Program of China (Grant No. 2017YFA0303302 and 2018YFA0305601), National Natural Science Foundation of China (11934005, 61322407, 11874116, 61674040), the Science and Technology Commission of Shanghai (Grant No. 19511120500), the Shanghai Municipal Science and Technology Major Project (Grant No. 2019SHZDZX01), and the Program of Shanghai Academic/Technology Research Leader (Grant No. 20XD1400200). E.Z. acknowledges support from China Postdoctoral Innovative Talents Support Program (Grant No. BX20190085) and China Postdoctoral Science Foundation (Grant No. 2019M661331). Part of the sample fabrication was performed at Fudan Nano-fabrication Laboratory. We thank Xuejian Gao from Prof. Kam Tuen Law's group from Hong Kong University of Science and Technology for helpful discussion on the Josephson interference mechanism.

**Author contributions**
F.X. conceived the ideas and supervised the overall research. Z.J. and Y.Z. synthesized high-quality $NbSe_2$ bulk samples. X.S. synthesized high-quality $Cr_2Ge_2Te_6$ bulk samples.

L.A., Z.J. and Y.Z. fabricated the nanodevices. L.A., E.Z. and C.H. performed the PPMS measurements and data analysis. X.X. and Y.Y. provided the curve fitting. S.L., Z.L. and P.L. carried out the SEM and AFM measurements. X.K. and Z.H. provided in-depth discussions of vdW physics. L.A. and F.X. wrote the paper with assistance from all other co-authors.**References**

1. Novoselov, K. S. *et al.* Electric Field Effect in Atomically Thin Carbon Films. *Science* **306**, 666 (2004).
2. Geim, A. K. & Grigorieva, I. V. Van der Waals heterostructures. *Nature* **499**, 419–425 (2013).
3. Liu, Y. *et al.* Van der Waals heterostructures and devices. *Nat. Rev. Mater.* **1**, 16042 (2016).
4. Lee, C.-H. *et al.* Atomically thin p–n junctions with van der Waals heterointerfaces. *Nat. Nanotechnol.* **9**, 676–681 (2014).
5. Avsar, A. *et al.* Spin–orbit proximity effect in graphene. *Nat. Commun.* **5**, 4875 (2014).
6. Zhong, D. *et al.* Van der Waals engineering of ferromagnetic semiconductor heterostructures for spin and valleytronics. *Sci. Adv.* **3**, e1603113 (2017).
7. Ben fez, L. A. *et al.* Tunable room-temperature spin galvanic and spin Hall effects in van der Waals heterostructures. *Nat. Mater.* **19**, 170–175 (2020).
8. Choi, K., Lee, Y. T. & Im, S. Two-dimensional van der Waals nanosheet devices for future electronics and photonics. *Nano Today* **11**, 626–643 (2016).
9. Fert, A. Nobel Lecture: Origin, development, and future of spintronics. *Rev. Mod. Phys.* **80**, 1517–1530 (2008).
10. Staley, N. E. *et al.* Electric field effect on superconductivity in atomically thin flakes of $NbSe_2$. *Phys. Rev. B* **80**, 184505 (2009).
11. Huang, B. *et al.* Layer-dependent ferromagnetism in a van der Waals crystal down to the monolayer limit. *Nature* **546**, 270–273 (2017).
12. Gong, C. *et al.* Discovery of intrinsic ferromagnetism in two-dimensional van der Waals crystals. *Nature* **546**, 265–269 (2017).
13. Deng, Y. *et al.* Gate-tunable room-temperature ferromagnetism in two-dimensional $Fe_3GeTe_2$. *Nature* **563**, 94–99 (2018).
14. Gingrich, E. C. *et al.* Controllable 0–π Josephson junctions containing a ferromagnetic spin valve. *Nat. Phys.* **12**, 564–567 (2016).
15. Baek, B., Rippard, W. H., Benz, S. P., Russek, S. E. & Dresselhaus, P. D. Hybrid superconducting-magnetic memory device using competing order parameters. *Nat. Commun.* **5**, (2014).
16. Buzdin, A. I. Proximity effects in superconductor-ferromagnet heterostructures. *Rev. Mod. Phys.* **77**, 935–976 (2005).
17. Ryazanov, V. V. *et al.* Coupling of Two Superconductors through a Ferromagnet: Evidence for a π Junction. *Phys. Rev. Lett.* **86**, 2427–2430 (2001).
18. Kontos, T. *et al.* Josephson Junction through a Thin Ferromagnetic Layer: Negative Coupling. *Phys. Rev. Lett.* **89**, 137007 (2002).
19. Kemmler, M. *et al.* Magnetic interference patterns in 0-π9

# Figures and Captions

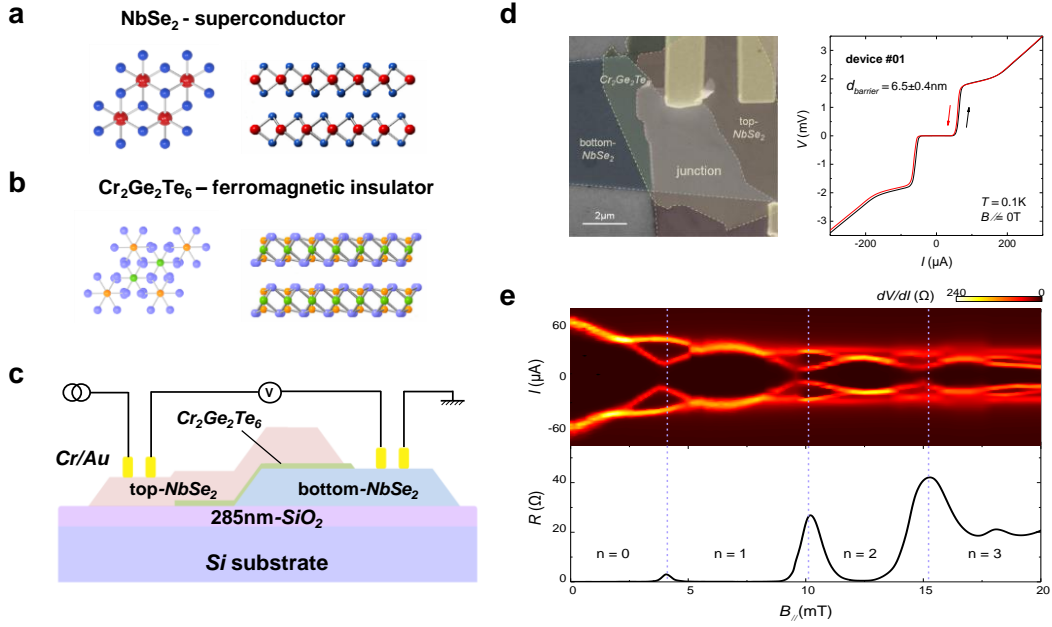

**Figure 1 | vdW Josephson junctions based on 2D superconductor and ferromagnetic materials. (a), (b)** Top and side views of the layered structures of superconductor NbSe$_2$ (Nb, red and Se, blue) and ferromagnetic insulator Cr$_2$Ge$_2$Te$_6$ (Cr, green, Ge, orange and Te, purple), while the easy axis of the latter one is out-of-plane. **(c)** Schematic illustration of the junction with the measurement configuration. A four-terminal device on the laminated structure is performed to exclude contact resistance, and the supercurrent flows across the interface. **(d)** False-color scanning electron microscope (SEM) image (left) of device #01 (scale bar, 2μm) and the trilayers are labeled by the dash lines respectively. The junction area is ~ 14.2 μm$^2$. The right figure in (d) is a typical current-voltage (*IV*) curve of JJ (device #01) at 0.1 K under zero magnetic field with the trap (black) and retrap (red) situation, and thickness of the barrier is ~ 6.5 nm measured by atomic force microscope (AFM). **(e)** Contour plot of differential resistance (*dV/dI*) as a function of bias current (*I*) and in-plane magnetic field, as well as periodically oscillatory in-plane magnetoresistance, showing a Fraunhofer pattern (of device #01) at 0.1 K corresponding to integer flux quantum Φ$_0$. Besides, higher-order oscillations due to Fiske resonance indicate a highly transparent junction surface.

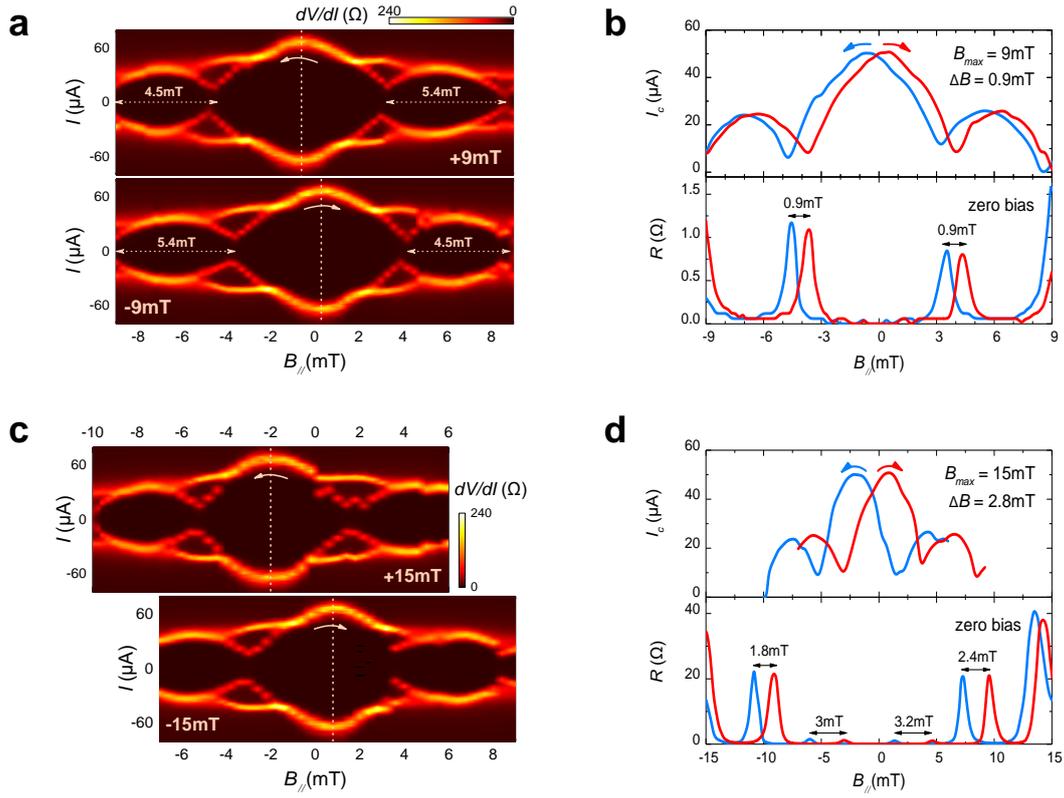

**Figure 2 | Hysteretic Fraunhofer patterns of the unsaturated barrier under in-plane magnetic field.** (a) Fraunhofer-like oscillation patterns of device #01 when the magnetic field sweeps from positive to negative (upper) and inverse (beneath), starting from ±9 mT. The yellow dashed lines mark out the central maximal of critical current, with respect to distinct field positions. (b) Hysteresis of measured critical current ($I_c$) and the corresponding junction resistance ($R$) in opposite field sweep directions shown in left figure (red: negative to positive, blue: back to negative, the followings are the same), where the effective field seen by the junction has an additional value of 0.9 mT originating from the remnant barrier moment. (c) Similar patterns of the same device above when the field starts from ±15 mT. (d) Hysteretic $I_c$ and $R$ for the maximum field applied of 15 mT, which gives a larger contribution to the total flux in the junction of 2.8 mT for a magnetic barrier whose magnetization is still unsaturated.

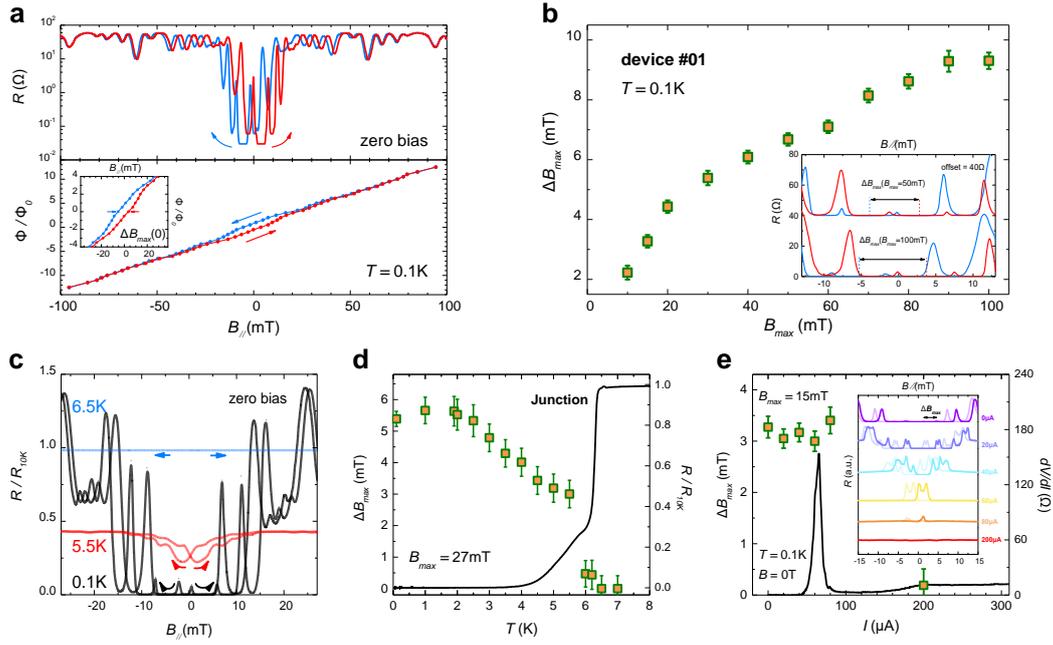

**Figure 3 | Evolution of the barrier magnetization in Josephson junctions. (a)** Hysteretic in-plane magnetoresistance (log-plot) of device #01 with maximum applied field ($B_{max}$) of 100 mT at 0.1 K (upper), in which the coincidence over ±75 mT reveals a saturated barrier magnetization. The field dependence of trapped flux in the junction depicts the magnetization curve of the barrier itself (beneath). The largest hysteresis ($\Delta B_{max}$) estimated from the central position of its zero-order minimum of $R$ is 9.3 mT. **(b)** Dependence of $\Delta B_{max}$ with regard to $B_{max}$, performing a continuous increase and finally converging at higher field. Right inset: comparison of the hysteretic $R$ and $\Delta B_{max}$ with a $B_{max}$ of 50 and 100 mT, respectively. The offset of the former is 40 Ω for clarity. **(c)** In-plane $R$ compared between the superconducting state and normal state (black for 0.1 K, red for 5.5 K and blue for 6.5 K). Magnetic hysteresis versus field sweep direction below the $T_c$ of NbSe$_2$ corresponds to the remanent magnetization of barrier Cr$_2$Ge$_2$Te$_6$, however, it disappears above $T_c$, suggesting that the magnitude of superconducting interference with temperature reflects in the size of hysteresis. **(d)** Temperature dependence of normalized junction $R$ (solid dark line) and $\Delta B_{max}$ (with a $B_{max}$ of 27 mT), where the value of the latter one remains almost unchanged below 2 K and gradually decreases up to $T_c$ of NbSe$_2$ (6.4 K). **(e)** $\Delta B_{max}$ (with a $B_{max}$ of 15 mT) at different d.c. bias current along with zero field $dV/dI$ curve (solid dark line) at 0.1 K, which is identical (within range of error) below 100 μA but virtually zero at high bias indicating the disappearance of hysteresis. Right inset: hysteretic $R$ at 0 to 200 μA bias current from superconducting to the normal state (stacked for clarity).

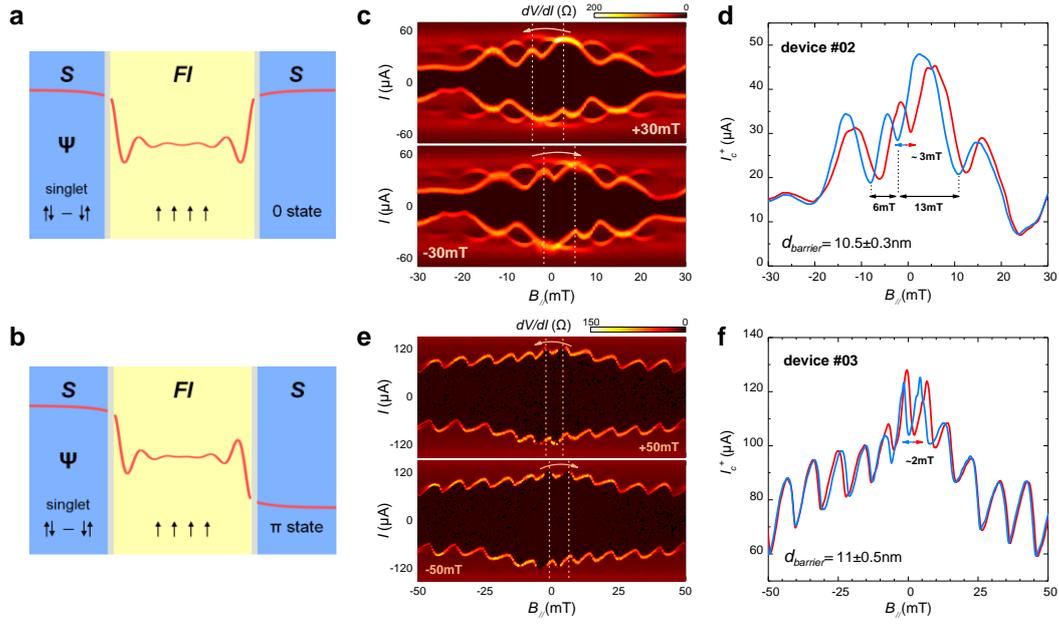

**Figure 4 | Evidence of π phase coupling through the magnetic Josephson junction.**
**(a), (b)** Illustrations of Josephson coupling situations with a thin barrier. The amplitude of Cooper pair wavefunctions Ψ is damping and oscillating in the ferromagnet with two phases of the ground state at 0 or π. **(c)** 0-π junction like oscillation patterns of a sample whose barrier thickness is ~ 10.5 nm (device #02) with the magnetic field starts from ±30 mT at 0.1 K. The yellow dashed lines mark the double peaks positions (at positive bias) around the central minimum of $I_c$ to depict a 0-π junction consisting of segments of 0 and π phases. **(d)** Hysteresis of $I_c^+$ of device #02 (blue for downward and red for upward). The value is estimated from the valley in the center that gives the result of 3 mT. **(e)** Another 0-π junction signatures of a sample with ~ 11 nm thick barrier (device #03), the field starting from ±50 mT at 0.1 K. **(f)** Hysteretic $I_c^+$ of device #03. The value of hysteresis is around 2 mT.

# Supplementary Information for

# Van der Waals Ferromagnetic Josephson Junctions

Linfeng Ai[1,2,†], Enze Zhang[1,2,†], Ce Huang[1,2,†], Xiaoyi Xie[1,2], Yunkun Yang[1,2], Zehao Jia[1,2], Yuda Zhang[1,2], Shanshan Liu[1,2], Zihan Li[1,2], Pengliang Leng[1,2], Xingdan Sun[3], Xufeng Kou[4], Zheng Han[3], Faxian Xiu[1,2,5,6*]

[1] State Key Laboratory of Surface Physics and Department of Physics, Fudan University, Shanghai 200433, China
[2] Institute for Nanoelectronic Devices and Quantum Computing, Fudan University, Shanghai 200433, China
[3] Shenyang National Laboratory for Materials Science, Institute of Metal Research, Chinese Academy of Sciences, Shenyang 110016, China
[4] School of Information Science and Technology, ShanghaiTech University, Shanghai 201210, China
[5] Collaborative Innovation Center of Advanced Microstructures, Nanjing 210093, China
[6] Shanghai Research Center for Quantum Sciences, Shanghai 201315, China

[†]These authors contributed equally to this work.
[*]Correspondence and requests for materials should be addressed to F. X. (E-mail: Faxian@fudan.edu.cn).

**Contents:**
I.   Additional results of device #01
II.  Illustrations of discontinuity and irreversibility in $I_c$ patterns
IV.  Excess Josephson current in thicker barrier device
V.   Thickness-tuned crossover from Josephson coupling to quantum tunneling
VI.  References

## I. Additional results of device #01

Additional transport properties of device #01 are shown here. Figure S1a is a zoom-in picture to show the difference between the trap and retrap situations in this sample's *IV* curves, which are integrated from the d*V/dI* curve plotted in the inset. Figure S1b is the temperature dependence of junction resistance (black) from 0.05 K to 300 K, and the calculated residual resistance ratio (RRR) at 10 K compared to 300 K is 0.15, which is of a similar magnitude to pure $NbSe_2$ (~ 0.12). A previous study in NbN/GdN/NbN JJ[1] showed that its junction behavior at high temperature is typical of a semiconducting barrier and then it turns into a continuous decrease in resistance below $T_{Curie}$ of GdN, due to the exchange splitting of the tunnel barrier with reduction of effective barrier height for one spin sign. However, the decrease of resistance in our device is dominated by $NbSe_2$ themselves. We have summarized the data of all devices, for each performs a metallic behavior before its superconducting transition temperature. Figure S1c presents the *IV* curves measured at the different magnetic fields as labeled in the inset, and the arrows indicate two current steps at bias voltage because of the Fiske resonance. In device #01, we can see a second-order Fiske step in which the positions of the 2$^{nd}$ maximum in $I_c$ correspond to the 1$^{st}$ minimum. Besides, a third-order Fiske step occasionally appears in device #02, as shown in the main text Figure 4c.

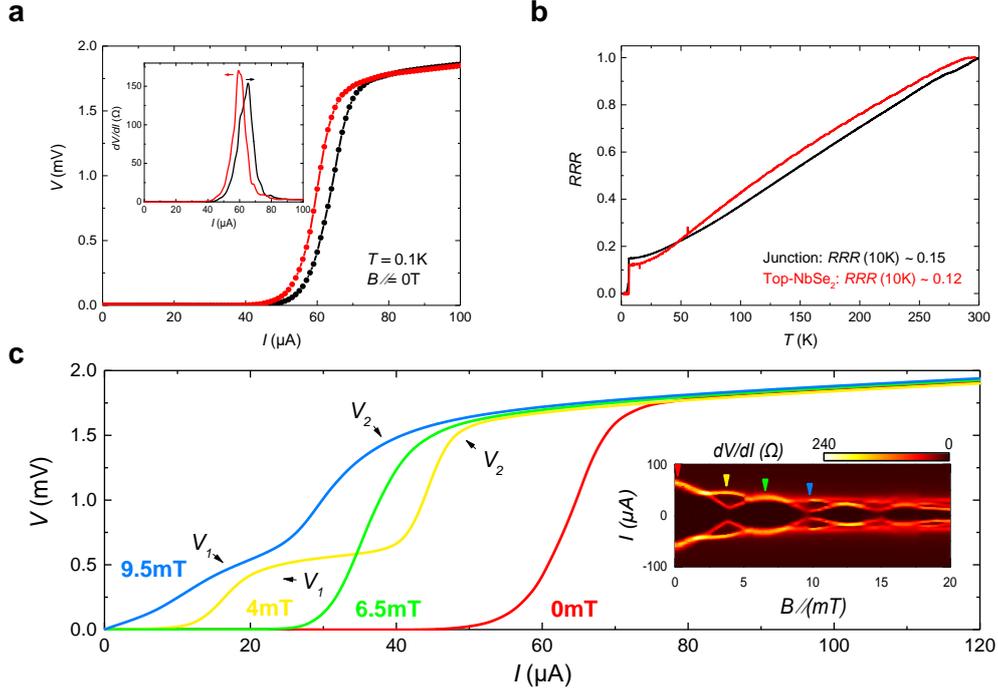

**Figure S1 | Additional results of device #01 as a JJ. (a)** *IV* curves of device #01 under zero field at 0.1 K with the trap (black) and red (retrap) configurations. The ratio of $I_r / I_c$ is 0.928 (the values of $I_c$ is extracted by 0.5$R_n$ from the *dV/dI* curve in the inset). **(b)** Temperature dependence of resistance of the junction (black) and $NbSe_2$ (red) from 0.05 K to 300 K, which shows a metallic barrier behavior above $T_c$. **(c)** *IV* curves measured at the different magnetic fields to reveal the Fiske steps. The inset shows a Fraunhofer pattern of device #01 with a second-order oscillation labeled

by the colored arrows.

A detailed response to the parallel magnetic field in device #01 is shown in the next parts. Figure S2a shows the magnetoresistance results measured at 1.9 K. For the junction, its resistance increases below 60 mT and eventually decreases up to 9 T when spins in the intermediate $Cr_2Ge_2Te_6$ were totally aligned to the field direction. In the meanwhile, the Ising superconductors $NbSe_2$ with a large parallel critical field (at the top and bottom) were still in the superconducting state with zero resistance, suggesting that the entire contribution of junction resistance comes from the tunneling barrier. In Figure S2b we show that the illustrations of barrier moment as a sequence of applied field split into three stages from the junction resistance, which corresponds to a complete hysteresis magnetization loop of magnetic barriers.

During the experiments, we have used two measurement systems to carry out the transport data in different temperature ranges, therefore a comparison between them is needed. From the arrows in Figure S2c we can see the central positions of two sweep branches are slightly shifted, meanwhile, the oscillations at 0.1 K are evident, compared to 1.9 K due to less thermal perturbations. Nevertheless, the magnitude of hysteresis is almost the same at these two temperatures (for one starts from 30 mT and the other from 27 mT), which also proves the validity that we use $\Delta B_{max}$ as an index for estimation from 7 K to 0.1 K.

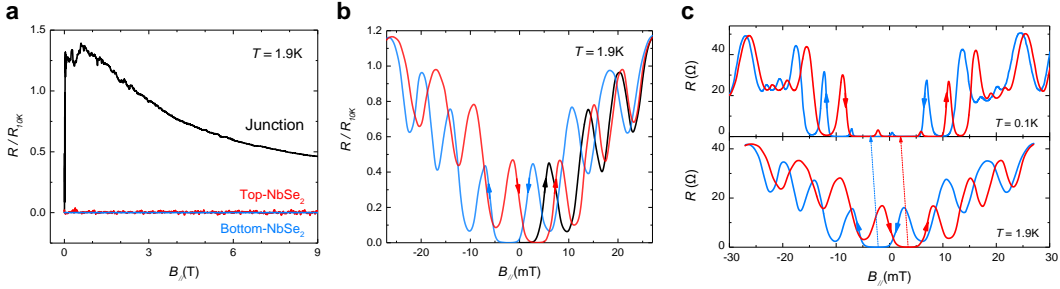

**Figure S2 | Additional magnetic response of device #01 at low temperature.** **(a)** Magnetoresistance of the junction (black) and pure superconductor layers (red for the top and blue for the bottom) versus parallel magnetic field up to 9 T. **(b)** Oscillatory resistance patterns of the junction at 1.9 K. The black line is measured after zero-field cooling to induce an initial magnetization of the barrier, and the blue and red curves are responses in downward and upward directions of field sweeping, respectively. **(c)** Comparison of hysteretic resistance patterns between the two measurement systems in different temperature ranges (0.1 K in a dilution refrigerator and 1.9 K in a PPMS).

Since the hysteresis introduced by the magnetic barrier in JJs can be controlled by $B_{max}$, we show a stacked plot of hysteretic junction resistance with $B_{max}$ ranging from 20 mT to 100 mT in Figure S3a. The coincidence of the opposite direction sweeping curves over 75 mT shown in the bottom indicates a saturated magnetized state, in contrast to smaller $B_{max}$ situations that the shift between the two curves arises immediately as the field direction turns back. Consequently, with a step-increase of $B_{max}$, the central lobe (equivalent to zero-flux state in the junction) moves as guided by arrows in Figure S3a, and the variation of its peak positions at downward or upward sweeping is presented in Figure S3b (1L represents the 1$^{st}$ maximum resistance at the left side of the central lobe while 1R at right side). Figure S3c depicts the separations between the two sweeping branches at different $B_{max}$, from which we can clearly see a rapid increase in measured hysteresis since the initial magnetization, and then it gradually converges above 75 mT, the same as the conclusion reached from Figure S3a.

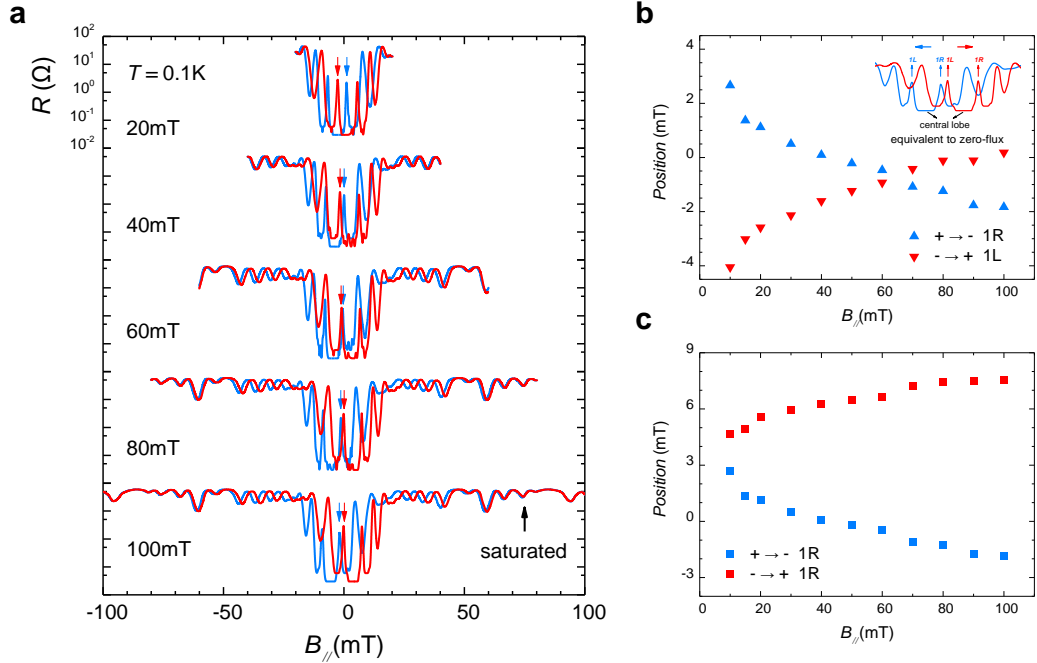

**Figure S3 | Hysteretic junction resistance at different $B_{max}$ of device #01. (a)** Log-plot of junction magnetoresistance measured with $B_{max}$ ranging from 20 mT to 100 mT at 0.1 K (blue curves for downward while red ones for upward). The barrier moment becomes saturated over 75 mT. **(b)** The variation of the two sweeping branches as shown in Figure S3a. Inset: illustrations of the central lobe (equivalent to zero-flux state in the junction) and peak 1L/1R in downward (blue) and upward direction (red), respectively. **(c)** The barrier's hysteresis evaluated from the corresponding peak 1R positions both for downward and upward direction.

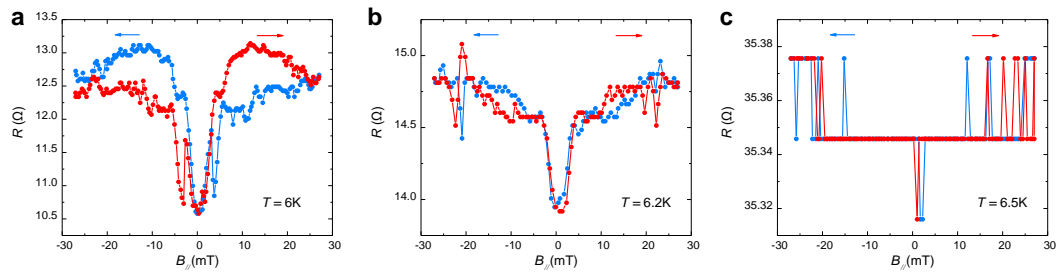

**Figure S4 | Junction magnetoresistance of device #01 near $T_c$. (a), (b), (c)** Junction resistance (blue for downward and red for upward) measured at 6 K, 6.2 K and 6.5 K. The discrepancies in the center positions of the curves tend to be illegible as the temperature approaches $T_c$, which causes a sudden drop of $\Delta B_{max}$ value as the data

shown in the main text Figure 3d. When the temperature is above $T_c$ ( ~ 6.4K), measured hysteresis is almost zero within the error of field resolution.

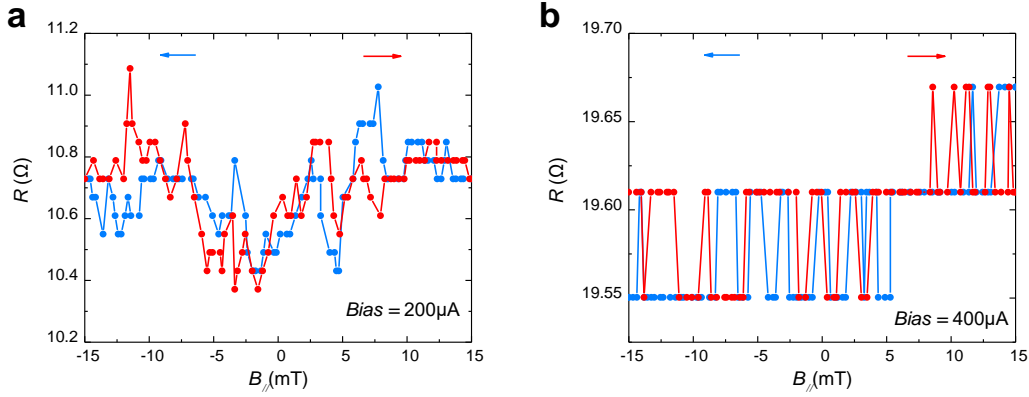

**Figure S5 | Junction magnetoresistance of device #01 at high *d.c.* bias. (a), (b)** Junction resistance measured at the *d.c.* bias of 200 μA and 400 μA. The deviations here are also practically unidentifiable at a high bias when the superconductor mainly contributes to the junction resistance.

**Supplementary Section II. Excess Josephson current in thicker barrier device**

For a thinner barrier like device #01 (~ 6.5nm), the tunneling behavior shows a good coincidence to the AB relation, whose general form implies an invariance of $I_c$ and the normal state resistance ($R_n$) depends on the measured superconducting energy gap. Note that the $R_n$ is ~ 10.6 Ω obtained from the linear part in its *IV* curves and the fitting results are shown in Figure S6a. The calculated $I_cR_n$ product ($V_c$) at zero temperature is ~ 0.51 mV, which the magnitude is much smaller than the theoretical value of $\pi\Delta_0/2e$ = 1.57 mV (bulk $NbSe_2$ energy gap: $\Delta_0$ ~ 1 meV). The strong suppression shows that the tunneling probability of pairs in a magnetic barrier is lower than that for single electrons.

However, in a thicker barrier device #03 (~ 11nm), we find an apparent excess Josephson current substantially deviated from the AB relation at low temperature. The data and fitting results (with a data range of 4.6 K to 6 K) are shown in Figure S6b. The $I_cR_n$ product obtained at the base temperature of 0.5 K is 1.43 mV, exceeding the diffusive limit from the AB relation fit at zero temperature (0.84 mV). We tend to attribute it to a possible multi-domain structure in few-layer $Cr_2Ge_2Te_6$ especially for thicker layers, and the introduced magnetic inhomogeneity by the relative size of domain walls implies a reduction of exchange energy in the ferromagnets, which accounts for the enhancement.

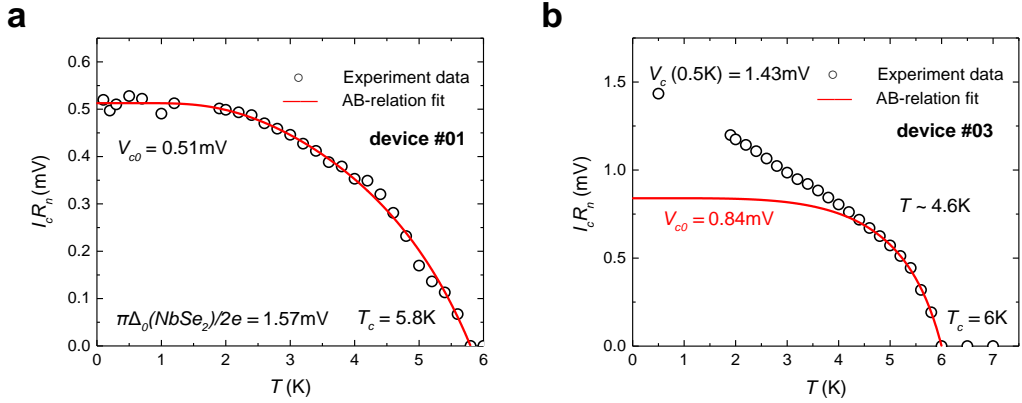

**Figure S6 | Multi-domains induced excess Josephson current in the thick barrier device.** (**a**) Temperature dependence of $I_cR_n$ ($V_c$) product of device #01 (black open circles) with a 6.5 nm barrier, and the red solid line is the fit to AB relation for an insulating barrier which gives $V_{c0}$ = 0.51 mV. The theoretical value is ~ 1.57 mV for NbSe$_2$ JJ. (**b**) Temperature dependence of $I_cR_n$ ($V_c$) product of device #03 with an 11 nm barrier and the largest $V_c$ acquired at 0.5 K is 1.43 mV. The red solid lines is an AB relation fit from 4.6 K to 6 K which gives $V_{c0}$ = 0.84 mV. The deviation between measured data and the fit is originated from the multi-domain structure in few-layer Cr$_2$Ge$_2$Te$_6$ especially for thicker samples, where the magnetic inhomogeneity can induce an equivalent reduction of exchange energy $E_{ex}$.

## Supplementary Section III. Illustrations of discontinuity and irreversibility in $I_c$ patterns

The breakpoints that appeared in $I_c$ patterns arise with the progressive magnetization of a highly polarized barrier, manifested as the abrupt jumps of $I_c$ evidently at higher field. In Figure S7, we first show the breakpoints measured in device #01 and the $dV/dI$ curves on both sides of the breakpoint near 25 mT. The conspicuous discontinuities above 23 mT are in sharp contrast to the patterns at the lower field region (the single step of the applied field is 0.5 mT in the whole process). Nevertheless, the results in the higher field become less recognizable because of the non-zero junction residual resistance at zero bias, thus we cannot get a valid $I_c$ value there.

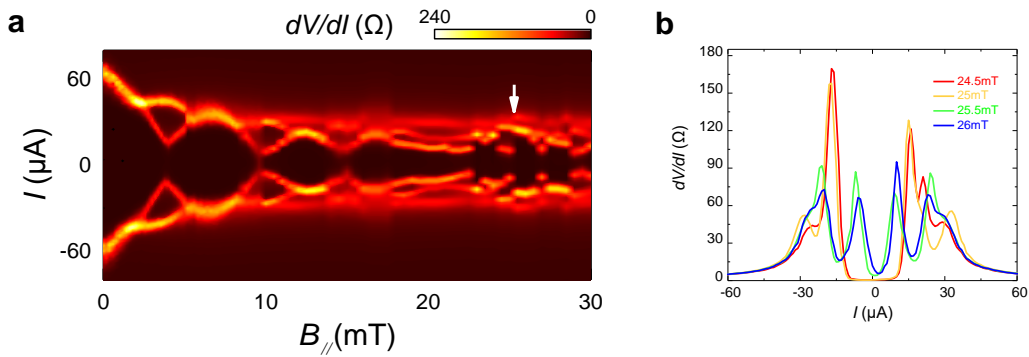

**Figure S7 | Illustrations of discontinuity in device #01.** (**a**) The Fraunhofer pattern

of device #01 initially magnetized to 30 mT at 0.1 K, where the discontinuity in $I_c$ starts from 23 mT. **(b)** $dV/dI$ curves under different magnetic fields near the breakpoint at 25 mT (indicated by white arrow in Figure S7a). The $I_c$ measured at 24.5 and 25 mT are almost identical (10.5 μA), but then encounter a sudden drop to almost zero as measured at 25.5 and 26 mT. The step of field resolution is 0.5 mT.

In device #02, the zero resistance state persists till -140 mT, in case that we can observe its first appearance of breakpoints at around -90 mT (Figure S8a). When the field reverses back to zero, the breakpoints still maintain their dominance even in the range where $I_c$ has previously varied continuously (Figure S8b). Apart from the interlayer coupling in $Cr_2Ge_2Te_6$ barrier which prevents the flip back of spins as we have discussed in the main text, the assumptions on multi-domains state possessing discrepant coercivity in the junction region may also be suggested to explain the step-like behaviors in $I_c$.

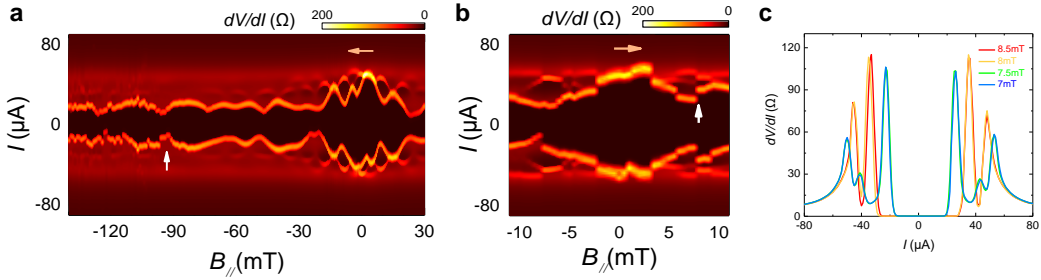

**Figure S8 | Illustrations of discontinuity and irreversibility in device #02. (a)** Fraunhofer-like oscillation patterns of device #02 ranging from 30 mT to -140 mT at 0.1 K, where the breakpoints in $I_c$ firstly appear at around -90 mT (marked by the white arrow). **(b)** Remeasured $I_c$ patterns near zero magnetic field after reversing back from -140 mT, presents pronounced irreversibility. **(c)** $dV/dI$ curves under different magnetic fields near the breakpoint at 8 mT, which is pointed by the white arrow in (b).

## Supplementary Section V. Thickness-tuned crossover from Josephson coupling to quantum tunneling

In addition to the JJ devices with relatively thin $Cr_2Ge_2Te_6$ as barriers, we have also investigated the crossover from the Josephson coupling state to the quantum tunneling dominant situation tuned by the magnetic barrier thickness. For a thick insulating layer, the Cooper pair wavefunction penetrating the JJ has damped to zero in the middle of the barrier that cannot provide a stable phase coherence, thus the transport conductance of the junction here is dominant by quantum tunneling effect through single-electron channels[2] (Figure S9a).

In the left inset of Figure S9b, we first display the normalized differential tunneling conductance ($G_s / G_n$) at 2 K for a 14.5±0.4 nm thick $Cr_2Ge_2Te_6$ junction device #04. The separation between the peaks corresponds to the superconducting gap of bi-lateral few-layer $NbSe_2$ (20 ~ 40 nm), whose properties are similar to the bulk $NbSe_2$. For the quantitative analysis of the tunneling spectra, we use the Blonder-Tinkham-Klapwijk (BTK) model[3] for SIS tunnel junctions with arbitrary barrier strength. Here, the temperature dependence of the gap (Figure S9 b) is described by the BCS theory, and the extracted zero-temperature gap value is $2\Delta_0 = 1.98$ meV quantitatively consistent with the value 3.53 $k_B T_c$ ($k_B$ is the Boltzmann constant and $T_c \sim 6.7$ K is consistent with the resistance measurements). Dimensionless effective barrier strength Z is 0.87 at 2 K, which is comparable to other vdW tunneling materials[4,5].

The magnetic response of the tunnel junction versus the in-plane field is shown in Figure S9c, at zero bias voltage. The resistance is normalized to the value at different temperatures in the absence of field. The hysteresis shrinks in resistance amplitude and magnetic field range with increasing temperature, albeit it fully disappears above $T_c$, similar to what we have observed in JJ devices. It's also noteworthy that the junction resistance near zero field is smaller than that in the higher field, and such a dip possibly accounts for a lower effective barrier height corresponding to zero magnetization compared to the fully-polarized state, which may also be influenced by the two spin-mixed S/F heterointerfaces to generate an anisotropic magnetoresistance[6,7].

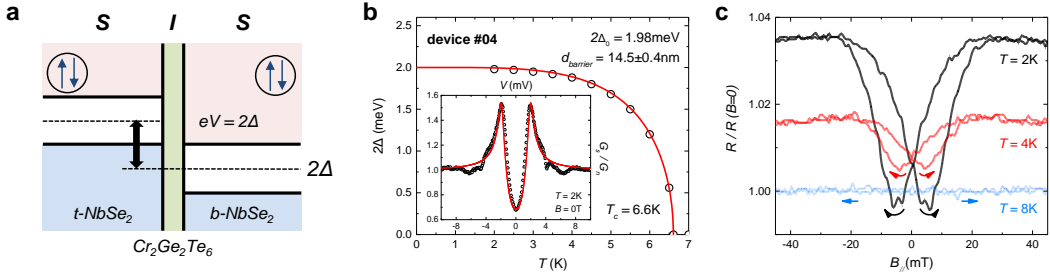

**Figure S9 | Quantum tunneling results of thick barrier devices.** (a) Schematic views of the energy diagram of the SIS tunnel junction with the same superconductors, showing that the conductance as a function of bias voltage measures the BCS superconducting gap $2\Delta$. (c) Normalized differential tunneling conductance ($G_s/G_n$) at 2K under zero field for a thick barrier device #04 (~ 14.5nm), and the BTK fitting gives the zero temperature BCS gap $2\Delta_0$ for NbSe$_2$ of 1.98 meV (black opened circles for experiment data and red solid lines for fit curves). (d) In-plane magnetoresistance of device #04 at 2 K, 4 K and 8 K compared among the superconducting state and normal state.

Our results provide a barrier thickness-tuned crossover from quantum tunneling to Josephson coupling in NbSe$_2$/Cr$_2$Ge$_2$Te$_6$ devices, nevertheless, we cannot confirm that it is a continuous process. Both of the effective penetration length and surface impurities are required to accomplish an entire zero resistance state in highly transparent and defect-free junction, therefore we plot a summary of normalized residual resistance ($R_{2K}/R_{10K}$) varied with barrier thickness in the range of 6 - 15 nm. The results are divided into three regions: $R_{2K}/R_{10K} > 1$, $R_{2K}/R_{10K} = 0$ and intermediate step $0 < R_{2K}/R_{10K} < 1$. All the devices' resistance in the Josephson coupling regime (6 ~ 11nm) continuously decreases when cooling down, in sharp contrast to the tunneling devices (13 ~ 15nm), however, for the others (11 ~ 13nm) the junction resistance saturates to a finite non-zero value at low temperatures. A careful comparison of their temperature dependences is shown in Figure S10a. We suppose that the junction resistance is formed by a series-connection of two $NbSe_2$ electrodes and the non-superconducting $Cr_2Ge_2Te_6$ part, while the former is zero below $T_c$ and the latter is the main cause of residual resistance. It is suggested that the effective penetration length of $\Psi$ in the dirty limit[8] should be comparable to the critical thickness for Cooper pair tunneling in these magnetic JJs. Interfacial scattering and impurities will lead to a finite residual resistance value even in JJ samples of thinner barriers, as suppressions of penetration length which is equivalent to thicker barriers.

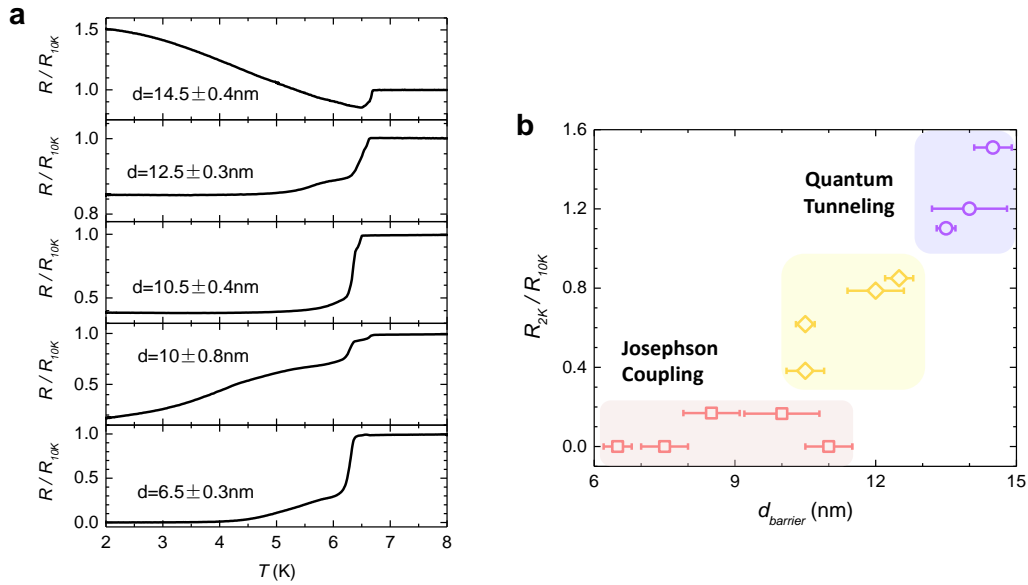

**Figure S10 | Summary of thickness-dependent tunneling process.** (**a**) Temperature dependence of different devices with a thickness of 6.5, 10, 10.5, 12.5 and 14.5 nm, respectively. The residual resistance measured at 2 K increases from 0 to over 1 suggesting the crossover from the Josephson coupling state to the quantum tunneling dominant state. (**b**) Normalized residual resistance at 2 K when tuning the $Cr_2Ge_2Te_6$ barrier thickness, in considerations of interface impurities and defects.